\documentclass[emulateapj,psfig,apjfonts]{emulateapj}
\submitted{{\it Accepted for publication in ApJ}}

\shortauthors{Walkowicz et al.}  \shorttitle{NUV Observations of M Dwarfs}
\begin{document}

\title{Characterizing the Near-UV Environment of M Dwarfs}

\author{
Lucianne M. Walkowicz\altaffilmark{1}, 
Christopher M. Johns-Krull\altaffilmark{2}, 
Suzanne L. Hawley\altaffilmark{1}
}

\altaffiltext{1}{Astronomy Department, University of Washington, Box 351580, Seattle, WA  98195}  
\altaffiltext{2}{Department of Physics and Astronomy, Rice University, 6100 Main Street, Houston, TX 77005}

\begin{abstract}
We report the results of our HST snapshot survey with the ACS HRC PR200L prism,
designed to measure the near-UV emission in a sample of nearby M dwarfs\footnote{Based on observations made with the NASA/ESA Hubble Space Telescope, obtained at the Space Telescope Science Institute, which is operated by the Association of Universities for Research in Astronomy, Inc., under NASA contract NAS 5-26555. These observations are associated with program 10525.}. 33
stars were observed, spanning the mass range from 0.1 - 0.6 solar masses
(T$_{eff}\sim$ 2200K - 4000K) where the UV energy distributions vary widely between
active and inactive stars. These observations provide much-needed constraints on models of the habitability zone
and the atmospheres of possible terrestrial planets orbiting M dwarf hosts, and
will be useful in refining the target selection for future space missions such as
TPF. We compare our data with a new generation of M dwarf atmospheric
models and discuss their implication for the chromospheric energy budget.
These NUV data will also be valuable in conjunction with existing optical, FUV
and X-ray data to explore unanswered questions regarding the dynamo generation
and magnetic heating in low-mass stars.
\end{abstract}


\section{Introduction}

M dwarfs are by far the most numerous of nearby stars, comprising roughly
75\% of the Galactic stellar population. The sheer number of M dwarfs in our
local solar neighborhood makes them attractive targets for planet-finding
missions, while their low masses make them a likely place for discovering
Earth-mass worlds in radial velocity surveys. However, these cool neighbors
also pose a unique set of problems for their resident planets, and so have been
largely dismissed in years past in favor of more tractable cases.

Recently, however, there has been renewed interest in M dwarfs as hosts for
habitable worlds. The case for M dwarf planets has been outlined in
\citet{t07}, bringing together studies by \citet{jo97}, \citet{jo03},
\citet{seg05} and others to propose that perhaps the case for habitability is
not so dire after all. As further timely motivation, \citet{udry} recently
announced the discovery of the lowest mass planet to date (M$_p \sim$
5M$_{\earth}$), orbiting in the habitable zone of a M3.5 dwarf, Gl 581.

The near-UV emission of M dwarfs remains an outstanding unknown, affecting both our understanding of the basic atmospheric structure in these stars and the potential habitability of attendant planets. Due to their cool effective temperatures, M dwarfs exhibit almost no continuum emission in the near-UV. However, a large
fraction of these stars possess surface magnetic fields, which give rise to a
steady-state, hot outer atmosphere and transient events such as starspots and
flares. Non-radiative heating processes transfer energy from the magnetic field
into the stellar atmosphere to power these phenomena, producing significant UV
output in the form of chromospheric (T$\sim 10^4$K), transition region (T$\sim
10^5$K) and coronal (T$\sim 10^6$K) emission. Chromospheric near-UV emission can have critical consequences for planetary atmospheres, as it directly
impacts the habitability of a neighboring planet through its role in the
production of ozone, as well as its implications for long-term survival (and
thus detectability) of biomarker gases (such as N$_2$O and CH$_4$). Preliminary studies of M dwarf habitability zones indicate
that the typical chromospheric UV flux from an active star will produce observable ozone layers,
comparable to that of Earth, on terrestrial planets around M dwarfs (see
\citet{seg05}, Figure 3). However, models of planetary atmospheres around relatively inactive M stars are presently based on stellar model photosphere predictions due to the
lack of near-UV observational data.

The data described in this paper also provide crucial constraints for new model atmospheres, which may be used to address a number of outstanding problems in stellar astrophysics. For example, optical (H$\alpha$) observations of nearby stars (\citet{ja74}; \citet{rhg95}; \citet{hgr96}; \citet{gi00};\citet{we04}) show that the active fraction (as measured by H$\alpha$ emission with equivalent width $>$ 1$\mbox{\AA}$) increases dramatically from types M0 to M6, peaks near spectral type M7-M8 and declines thereafter. The
strength of activity, as measured by the ratio of the luminosity in H$\alpha$
compared to the bolometric luminosity ($L_{H\alpha}/L_{bol}$), also  declines
at late types (Burgasser et al. 2002; Cruz \& Reid 2002; West et al. 2004). In addition,
it has been suggested that activity in ultracool dwarfs is confined primarily
to large flares, with little or no ``quiescent'' emission (\citet{ru00};
\citet{li95}; \citet{fgs00}). In contrast, STIS/MAMA observations of active
M7-M9 dwarfs by \citet{hjk03} showed that quiescent transition region (C IV)
emission is present at levels comparable to those seen in earlier active M
dwarfs.

The difficulty in interpreting these intriguing empirical data lies primarily in our
lack of a theory of magnetic dynamo field production and subsequent atmospheric
heating. At types later than about M3, stars  become fully convective and a
solar-type shell dynamo can no longer operate. The theory of $\alpha^2$ or
turbulent dynamos is not yet developed enough to provide predictions or
guidance regarding the efficacy  of magnetic field production.  Progress may be
made, however, on the atmospheric heating problem. It has not been established
whether the drop in active fraction and activity level seen at types later than
M7 is a result of the particular diagnostic we are using (H$\alpha$). If
magnetic heating is still present, but the atmospheric structure has changed in
these late type stars, emission may occur preferentially in other lines. In
particular, the relatively unobserved 1700-3000$\mbox{\AA}$ wavelength region contains
numerous strong emission lines important to the energy balance of the outer
atmosphere, including the resonance lines of Mg II and a host of Fe II emission
lines. These lines provide additional information on the atmospheric structure at
chromospheric and transition region temperatures, and therefore to the
underlying heating processes.

\section{Observations and Analysis}

Our sample comprises 107 nearby low mass dwarfs, extensively studied in the
Palomar/Michigan State University (PMSU) survey of nearby stars [\citet{rhg95},
\citet{hgr96}]. We have homogeneous optical spectra for nearly all of these stars,
which allowed us to choose a wide distribution of spectral types and
chromospheric activity strengths (the equivalent width of the H$\alpha$
line is used a proxy for chromospheric activity). Prior spectroscopic analysis
of these stars also allowed us to select our sample to exclude known
spectroscopic binaries. Roughly half of the stars in our
sample were previously observed with the International Ultraviolet Explorer (IUE) at
near-UV wavelengths (see discussion, Section 3.1). In addition, HST/STIS
spectra of the NUV and transition region FUV emission are available for
several of the most active stars in our sample.

The Hubble Space Telescope (HST) observed 29 M dwarfs and 2 late K dwarfs from
our sample using the Advanced Camera for Surveys High Resolution Channel (ACS
HRC) with the PR200L prism in Snapshot mode. Table 1 provides a summary of the observed
sample. Observations spanned the period of 3 August 2005 to 11 June 2006,
terminating with an electronics glitch in the HRC and subsequent safing of
the instrument. The observed sample ranges from spectral type K7 to M5
and includes both inactive and active stars. All but two of the observed
stars lie within 10 parsecs of the Sun.

With the failure of the STIS instrument in 2004, the PR200L prism became the
only spectroscopic element on board HST capable of near-UV spectroscopic obsevations. The PR200L has a highly nonlinear
dispersion, ranging from 5.3$\mbox{\AA}$ pix$^{-1}$ at 1800$\mbox{\AA}$ to 105$\mbox{\AA}$   pix$^{-1}$ at
3500$\mbox{\AA}$. By 5000$\mbox{\AA}$ the dispersion decreases to 563$\mbox{\AA}$   pix$^{-1}$, in essence
depositing all redwards flux onto just a few pixels of the CCD detector. The rapidly decreasing
dispersion towards longer wavelengths results in what has become known as the
``red pile-up'' \footnote{HST Instrument Science Report 0603}, making the PR200L
data problematic for the study of intrinsically red objects, where scattered light
from the red end can contaminate or drown the faint UV signal. Fortunately,
diffraction spikes from the red pileup lie at an angle to the dispersion
direction, causing the degree of contamination to lessen towards shorter
wavelengths. Taking advantage of this fact, we were able to choose an
extraction window that minimized the scattered light contamination.

Spectra were extracted in PyRAF using the aXe slitless spectroscopy reduction
package, which consists of several tasks designed to handle large format
spectroscopic slitless images. Observations for an individual star consist of a
``direct'' or undispersed image taken with the F330W filter, followed by a
dispersed image taken with the PR200L prism. In the reductions, the position of
the star on the direct image is used in combination with telescope position
information from the header to locate the spectrum on the subsequent dispersed
image and assign a wavelength to each pixel in the spectrum. aXe first
generates pixel extraction tables (PETs), which contain a spectral description
for each pixel in the spectrum. Pixels in the extraction window are then projected onto the spectral
trace in individual wavelength bins, and their signal is weighted by the
fractional area of the pixel that falls in a given bin. Flux calibration
follows using sensitivity curves for the relevant observing mode (here, the
HRC with PR200L prism). Although the potential for contamination by neighboring
objects in an image exists for all slitless spectroscopy (e.g. when dispersing
stars that lie in a crowded field), all of our targets were either solitary or well
separated from others in their images, so geometric contamination was not an
issue.

The reduction tasks require very little user interaction, most information
being supplied by STSDAS configuration files or drawn from the image headers
themselves. At the user's discretion, however, are the parameters of the
extraction window. As the diffraction spikes from the red pileup lie very close to the NUV
spectrum, a careful balance must be struck between including as much of the
already-meager NUV flux as possible, while excluding the influence of the
nearby artifact. As previously mentioned, the diffraction spikes lie at an
angle to the spectrum, merging with the spectrum at the red end. The artifact
and spectrum diverge towards shorter wavelengths, but if
a wide extraction window is used, part of the artifact may still be
included. The wavelength extent of contamination is therefore a function of the
width of the extraction window, where a wide window includes the artifact over
more of the spectrum than a narrow window does. In order to quantify the effect
of the artifact on the output spectrum, we experimented both with a variety of
extraction windows centered on the spectrum, and with extractions offset below
the spectrum (i.e. a slice across the diffraction spike) and above the spectrum (where
no artifact was present). After much experimentation, we determined a narrow
extraction window half-width of 1 pixel to be the best choice. To quantify the
uncertainty in our spectra, we extracted a spectrum from the background (offset
3 pixels above the spectrum) to examine the pixel-to-pixel
variation. Variations in the background spectra were used to compute the error
bars that are shown on all plots in this paper.

\section{Results}

Figure 1 shows the resulting near-UV spectrum for one of the most active stars in
our sample, AD Leo (dM3e, d=4.9pc). Note that the uncertainty in the flux increases towards longer
wavelengths as contamination from the red pileup becomes more apparent. In
Figure 1 we also compare prior IUE observations of AD Leo to our extracted ACS
spectrum. Plotted  at top is the IUE spectrum, taken with the low resolution
grating. At center we plot a simulated prism spectrum generated by using the
IUE spectrum as input to SLIM, the ACS slitless spectroscopy
simulator\footnote{http://www.stecf.org/instruments/ACSgrism/slim/index.php}. SLIM
uses an input spectrum of arbitrary resolution to simulate an ACS grism or prism image using a
specified configuration (in our case, the HRC with the PR200L prism). As SLIM
does not account for the artifacts introduced by the red pileup, the simulated
spectrum represents a ``best-case scenario'', devoid of contaminating scattered
light. The simulated spectrum shows three main features: a peak at 2800$\mbox{\AA}$,
attributable to the strong Mg II h and k lines, and two other peaks, at
2600$\mbox{\AA}$ and 2400$\mbox{\AA}$, both due to collections of strong Fe II lines
clustered around those wavelengths. These features are also seen in the
observed ACS spectrum for AD Leo, plotted at the bottom of the figure. The flux
in the Mg II feature at 2800$\mbox{\AA}$ matches almost exactly between the simulated and
observed spectra, and the fluxes in the two Fe II features at 2600$\mbox{\AA}$ and
2400$\mbox{\AA}$ are also quite comparable.

Figure 2 plots the observed NUV spectra for the entire sample, in approximate order of
strong to weak ultraviolet emission. For the most active stars, we clearly
resolve the main spectral features discussed in Figure 1. As activity
decreases, emission from the Mg II feature at 2800$\mbox{\AA}$ remains prominent, but
the Fe II features become less defined. Also, for a few objects, only the
emission peak near 2600$\mbox{\AA}$ is seen, without significant emission at 2800$\mbox{\AA}$ 
or 2400$\mbox{\AA}$ (see for example Gl 273, Gl 876 and Gl 682 in Figure 2b).

Although the prism data have very low resolution, we were able to extratct an estimate of the fluxes in the regions near the strong lines. The spectra were fit by four gaussians (centers held fixed at $\lambda$ = 2385$\mbox{\AA}$, 2560$\mbox{\AA}$, 2640$\mbox{\AA}$ and 2800$\mbox{\AA}$) and a second order polynomial. The area under the gaussians can then be used to characterize the flux due to emission features near those wavelengths. The result of one such fit is shown in Figure 3, and the fluxes measured for all spectra are listed in Table 2. If no flux is listed for a particular feature, that feature did not contribute significantly to the spectrum or the fit was unreliable.  

\subsection{Comparison to Existing Data}

Approximately half of our observed sample have existing IUE low resolution
spectra. Linsky et al. (1982) carried out an early pioneering study of
ultraviolet emission in active cool stars with IUE, while a later comparative
study of Mg II h and k emission with X-ray emission was undertaken by Mathioudakis \& Doyle (1989). However, the majority of stars observed with IUE are active, and the few lower activity objects with IUE spectra have relatively low S/N. 

In order to better compare the IUE and ACS observations, we used the existing IUE spectra to simulate ACS spectra with SLIM. As previously discussed, the results for AD Leo are shown in Figure 1. The three peaks in the AD Leo ACS spectrum at 2400$\mbox{\AA}$, 2600$\mbox{\AA}$ and 2800$\mbox{\AA}$ are recreated in the SLIM simulated spectrum, and the fluxes in the data and simulated spectrum are a good match within the uncertainties. Not all of the SLIM simulated spectra gave such good results-- many of the noisier IUE spectra produce accordingly noisier simulated spectra, so comparison between the simulated and observed fluxes is unreliable. As the best-exposed IUE spectra all had prominent Mg II emission, we chose to compare the Mg II flux between our ACS data and the IUE spectra as a test of our ability to measure accurate line fluxes. The flux in the ACS spectra due to the gaussian measured at $\lambda_c$ = 2800$\mbox{\AA}$ was taken to be representative of the Mg II flux. The IUE Mg II lines were fit with a single gaussian as the h \& k lines were blended in the low resolution data. Figure 4 shows IUE versus ACS Mg II surface fluxes, calculated using radii given in \citet{nlds} Table 4.1, and distances from trigonometric parallax measurements. The brightest and most active stars, for which Mg II emission was most prominent, show a good match between the ACS and IUE measurements. Fainter objects were probably more affected by scattered light in the ACS spectra, and therefore are measured as having more flux than the IUE spectra.  

Many of our stars have X-ray measurements from ROSAT. Figure 5 shows the
surface  ACS Mg II flux versus the ROSAT surface flux. Our results are very
similar to Figure 1 from \citet{md89}. Their flux-flux relation for dMe stars
(stars with H$\alpha$ in emission, represented as filled black circles) is overplotted as a dotted line. A slight positive offset exists
between the X-ray fluxes and the overplotted fit, as the \citet{md89} correlation used measurements from Einstein and EXOSAT rather than ROSAT [\citet{li93}], but otherwise our results are in excellent agreement.

Most of the stars in our sample also have measured H$\alpha$ equivalent widths,
originally reported in \citet{grh02} and included in Table 1 of this
paper. Using the $\chi$ factor of \citet{wa04}, we calculated the flux in
H$\alpha$ for each star. Figure 6 shows the computed H$\alpha$ flux versus the
log of the ACS Mg II flux, where negative flux values imply that the H$\alpha$
line is in absorption. 

Taken together, Figures 5 and 6 present a complex picture of chromospheric and
coronal emission. While the most ``active'' stars, such as AD Leo, have strong
emission in Mg II, X-rays and H$\alpha$, many of the stars possessing the
highest amounts of Mg II emission (e.g. Gl 205 and Gl 299) have relatively weak
X-ray emission and H$\alpha$ in absorption. These results indicate that there
is no simple relationship between the activity level (as indicated by
H$\alpha$) and where emission preferentially occurs, and the use of simple
scaling relations between these line fluxes may be misleading.

\subsection{Comparison to Models}

We compare the near-UV observational data with predictions from the radiative hydrodynamic models of Allred et al. (2005, 2006). 
The models use solutions to the 1-D equations of radiative hydrodynamics,
including non-LTE radiative transfer in H, He and Ca II, with flare heating
provided by an electron beam.  The results include predictions of atmospheric
structure (temperature, density profiles), velocity and line profiles at many
time steps during an episode of flare heating.  We used the results of the Allred
preflare and F10 models to investigate the Mg II and Fe II emission in and out
of flaring states (the F10 model used here represents an average mid-flare atmosphere
near a time step of $\sim$ 85 sec; see Allred et al. (2006)). In addition we used two other preflare models, scaled to be either cooler or hotter at a given column mass than the Allred preflare atmosphere. The four atmospheric models are designated here as PF, Flare, LPLT (low pressure low temperature), and HPHT (high pressure high temperature), for the preflare, flare, cooler and hotter atmospheres, respectively. The model atmospheres are shown in Figure 7.   

The Allred et al. models only include H, He and Ca II in the detailed non-LTE
treatment. However, thousands of other transitions (including those of Mg and
Fe) are included as background opacity in LTE. The solar models of Abbett \&
Hawley (1999; see section 2.4) compare the effects of including detailed
non-LTE calculations for Mg II or treating it as background opacity, and they
find that there is little change in the total cooling rate for these two
cases. Thus, including Mg and Fe in non-LTE is likely to make little difference
in the overall energy balance of the atmosphere. However, since Allred et
al. (2006) did not predict the Mg II or Fe II line profiles which comprise the majority of emission lines in our observed wavelength range,
we analyzed our atmosphere models with  the `RH' non-LTE
radiative transfer code described in Uitenbroek (2001). The RH code is based
on the Multi-level Accelerated Lambda (MALI) formalism of Rybicki \& Hummer
(1991, 1992),  which allows both bound-bound and bound-free radiative
transitions to overlap in wavelength. It also includes the effects of partial
redistribution for strong bound-bound transitions such as Mg II h and k. The
Mg II calculations for the PF and Flare atmospheres were originally reported in
\citet{yzcmi}. The non-LTE atomic level
populations were calculated for a 9 level hydrogen atom with continua, an 11
level Mg II ion with continua and a 143 level Fe II ion with continuum. Other
atomic transitions (of He, Ca, C, O, Si, Al, NaI and N) and molecular
transitions (of H$_2$+, C$_2$, N$_2$, O$_2$, CH,CO,CN,NH,NO OH and H$_2$0) are
included as background opacity in LTE.

While the PR200L resolution is clearly insufficient for comparison with the
detailed model line profiles, we can use these model spectra as inputs to SLIM,
giving us PR200L simulated low resolution spectra for Mg II and Fe II. In Figures 8-11, we show the model spectra for Mg II and Fe II for each atmosphere, the SLIM simulated ACS spectra for Mg II and Fe II separately, and the combined result. All fluxes have been scaled to a star with the radius and distance of AD Leo (R=0.41R$_{Sun}$, d=4.9pc) [\citet{scior}, \citet{nlds}] to enable direct comparison to the data. Overall, the shape of the model spectra are qualitatively similar to the observed spectra. One notable feature is that the cooler atmospheres (LPLT and PF) do not predict a strong emission peak at 2800$\mbox{\AA}$, and instead predict a trough centered there. The resulting spectra are then reminiscent of the ACS data for less active stars, such as Gl 889.1 and Gl 825 (shown in Figure 2b). In every case, our models overpredict the Fe II emission. The result for our ``Flare'' atmosphere shows the strong Mg II emission at 2800$\mbox{\AA}$ that we see in many spectra (especially of the active stars), but the Fe II emission predicted by this same atmosphere vastly overwhelms the Mg II peak, which is not observed. We reproduce the observed peaks in the ACS data due to Fe II, though emission at the shorter wavelengths is stronger than predicted. We note, however, that in the IUE simulated spectrum of AD Leo (Figure 1), the Fe II emission at 2400$\mbox{\AA}$ was stronger than observed in the ACS spectrum-- it therefore may be that the shortest wavelengths in the ACS spectra are not accurately measured as the signal is extremely faint.  
 
\subsection{Specific Cases}

We used a combination of the resulting model prism spectra to match the ACS
observations of Gl 876, Gl 825, Gl 889.1, Gl 273, and AD Leo. These specific
stars were selected on the basis of their relatively well exposed spectra and
variety of activity levels (as inferred by their H$\alpha$ equivalent
widths).In order to gain  insight into the structure of the active regions of
the chromosphere, we determine a ``fill factor'', or effective surface area
coverage of which model spectrum best fits these stars. In several cases, the observed data are best fit by a
combination of spectra from different atmospheres, indicating the probable
existence of multiple components in the chromospheres of these stars. The
best fits were determined by eye in each case, with preference toward matching
the flux between 2200$\mbox{\AA}$ and 2900$\mbox{\AA}$. Wavelengths longward of
2900$\mbox{\AA}$ lie nearer the red
pile-up, and therefore are more likely to be unreliable (for example, SLIM
simulations of both models and IUE data never recreate the deep dip at
2950$\mbox{\AA}$, which is seen in all the ACS spectra). 

\subsubsection{Gl 876 (dM4)}

Figure 12 shows the fit to the spectrum of Gl 876. We find that the PF atmosphere provides the best fit to the observations, with a fill factor of $\sim$1.3\%. Gl 876 is of particular interest for astrobiology, as it is host to 3 known extrasolar planets, including a $\sim$7.5M$_{\Earth}$ ``super Earth''. Gl 876 has an H$\alpha$ absorption line (EW$_{H\alpha}$ = $-$0.2$\mbox{\AA}$), indicating that it possesses a low-to-intermediate level of chromospheric activity (as compared with M dwarfs with H$\alpha$ lines in emission, which have a high level of chromospheric heating and are thus deemed ``active''). M dwarfs with very little activity and those with moderate activity cannot be distinguished on the basis of H$\alpha$ alone, as both weak and intermediate chromospheres show H$\alpha$ in absorption [\citet{cm85}]. However, information from other chromospheric lines (such as Ca II, Mg II and Fe II) may serve to break this degeneracy. 

With  H$\alpha$ in absorption and a spectral type of dM4,  Gl 876 is roughly equivalent in temperature and activity to the 3100K inactive star used in the \citet{seg05} biomarker study. \citet{seg05} used a model spectrum from \citet{all97} with no chromospheric flux for their cool, inactive M dwarf-- however, it is debatable whether such a star exists, as most M dwarfs show evidence of a chromosphere at some level [\citet{cm85}]. Low-to-intermediate activity stars, like Gl 876, are the true unknowns in studies of the effect of stellar flux on attendant planets-- while they may not have the strong chromspheric emission of very active stars like AD Leo, their near-UV emission may still be non-neglible. 

Scaling our spectrum of Gl 876 to the top-of-atmosphere flux for a planet in the habitable zone (semimajor axes given in \citet{seg05} Table 1), we find that the so-called ``inactive'' Gl 876 actually has an appreciable near-UV flux. In Figure 13, we plot our results for Gl 876 side by side with a similarly scaled spectrum of AD Leo (again using parameters from \citet{seg05} Table 1) and show that the expected near-UV flux at a planet in the habitable zone is actually comparable for both stars. Results for ozone production and methane chemistry in the atmosphere of a planet around Gl 876 would therefore be more in keeping with those given for AD Leo, rather than the predicted ``methane runaway'' in planets around quiescent M dwarfs [\citet{seg05}]. 
  
\subsubsection{Gl 273 (dM3.5)}

In Figure 14, we plot the resulting model fit to our ACS spectrum of Gl 273. Here, the data are best fit by a combination of our LPLT and PF atmospheres in roughly equal portion (0.6\% and 0.5\%, respectively. This fit gives an overall fill factor of $\sim$1\% for the active regions that produce the near-UV flux. 
 
\subsubsection{Gl 825 (dK7)}

Gl 825 is one of the hotter, more massive stars in our sample. We find that a
combination of the LPLT and PF atmosphere model spectra, at 3\%  and 4.5\% fill factor respectively, provides a good match to the majority of the spectrum (Figure 15). There is an excess of flux at $\sim$2750 that our model spectra do not reproduce, however, and increasing the fill factor to match this part of the spectrum then overestimates the Fe II flux at shorter wavelengths.

\subsubsection{Gl 889.1 (dM0)}

Gl 889.1 is best matched by a combination of the LPLT and HPHT atmospheres, at fill factors of 1.6\% and 0.05\% respectively. While the spectrum has a similar shape to that of Gl 825, Gl 889.1 has more short wavelength flux (from 2000 to 2200$\mbox{\AA}$) that is not well fit without the hotter HPHT component. The resulting fit is shown in Figure 16. 

\subsubsection{AD Leo (dM3e)}

Fitting our results for AD Leo is challenging, as the models which match the Mg II emission clearly overpredict emission from Fe II. As some Fe II also contributes flux to the region around 2800$\mbox{\AA}$, a much smaller fill factor is necessary to match the combined Fe II $+$ Mg II model spectra than what one would interpret from Mg II results alone. As AD Leo is a very active star, we first attempt to fit the data with our hottest ``Flare'' model (Figure 17, left panel). While a fill factor of 0.25\% reproduces roughly the correct amount of flux around 2800$\mbox{\AA}$ and at wavelengths shorter than 2200$\mbox{\AA}$, the Fe II emission at $\sim$ 2400$\mbox{\AA}$ and $\sim$2600$\mbox{\AA}$ is much too large. The HPHT atmosphere (also at a fill factor of 0.25\%; Figure 17, right panel) produces a much better match to these two Fe II features, but unfortunately does not account for any of the observed Mg II flux. Using the Mg II results for our Flare model alone, the emission near 2800$\mbox{\AA}$ is well matched with a fill factor of 1.6\%. While it is not physical to ignore the Fe II emission, the 1.6\% coverage required to
reproduce the Mg II flux provides an upper limit
estimate on the area filled by this hot ``Flare'' component.

\section{Conclusions and Future Work}

We obtained 31 ACS PR200L prism spectra of stars, covering spectral types K7 to M5, and including a variety of activity levels. The near-UV data presented here provide valuable constraints for both model stellar chromospheres and photochemical models of planetary atmospheres around M dwarfs. 

We measured Mg II fluxes for our spectra and compared our results with
complementary optical, near-UV and X-ray data. For those stars in our sample with
IUE data, we used the IUE Mg II flux as a check of our ACS prism spectra
measurements, and found them roughly consistent. We also
examined the relationship between Mg II emission and ROSAT measurements of the
X-ray flux. Our results agree with those of \citet{md89}, showing a well-defined
correlation for the most active dMe stars. Using existing H$\alpha$
measurements for our sample in conjunction with the $\chi$ factor of
\citet{wa04}, we computed H$\alpha$ fluxes and compared these to our
measured Mg II fluxes, finding a complex relationship between the optical and
near-UV emission. Together, the Mg II, X-ray and H$\alpha$ data discussed here present a
complex picture of the chromospheric and coronal emission in these stars.

Using the `RH' NLTE radiative transfer code, we calculated Mg II and Fe II
spectra for a suite of model atmospheres. The resulting model spectra were then
degraded to the resolution of the ACS prism data using SLIM, the ACS prism
spectra simulator. We compared our results with several of the observed spectra
to infer potential chromospheric structure and active region coverage, finding
that most stars could be matched by an active region area coverage of a few
percent of the stellar surface. This result is consistent with our findings in
\citet{yzcmi}, where we determined that $\sim$0.5\% of the visible stellar
surface was involved in the observed flares.  While in many cases the
qualitative shape of the model spectra matched the observations, Fe II was
overpredicted for the most active objects. The empirical results we find for Mg
II, X-ray and H$\alpha$ pose an interesting challenge for the models. The model atmospheres described in this paper will be used in future calculations of H$\alpha$ and Ca II K, which will then be compared with our results for the near-UV emission. 

In the case of Gl 876, a dM4 star that hosts three extrasolar planets, we scaled our observed spectra to match the expected flux at the distance of a planet in the habitable zone. We compared our results to a similar scaling of our AD Leo observation, and then discussed these results in the light of previous studies of M dwarf planet habitability. We find that our observation of what might be called an ``inactive'' star (on the basis of H$\alpha$, the usual proxy for activity in late type stars) actually shows a substantial amount of near-UV flux. Although Gl 876 is a lower luminosity and lower activity star than AD Leo, the non-negligible near-UV flux and proximity of the habitable zone conspire to make it a similar host for planets. The spectra described in this paper will be further used as input stellar flux for the Virtual Planet Laboratory's photochemical planetary atmosphere models.

\acknowledgements
The authors thank Pat Hartigan for his guidance with the reduction
process. LMW thanks Han Uitenbroek for the use of his RH code and Mats Carlsson
for his valuable input. LMW and SLH thank the Helen Riaboff Whiteley Center at
the University of Washington Friday Harbor Laboratories, in
whose phronistery this paper was completed. Support for program 10525 was
provided by NASA through a grant from the Space Telescope Science Institute, which is operated by the Association of
Universities for Research in Astronomy, Inc., under NASA contract NAS
5-26555. This research has made use of the aXe extraction software, produced by
ST-ECF, Garching, Germany.

\clearpage
\begin{deluxetable}{lllrrrcc}
\tablecaption{Observed Sample}  \tablenum{1}  \tablecolumns{8} \tablewidth{0pc}
\tabletypesize{\footnotesize}  
\tablehead{\colhead{HST Name} & \colhead{Name} & \colhead{NN} & \colhead{M$_v$\tablenotemark{a}} & \colhead{H$\alpha$\tablenotemark{a}} & \colhead{d (pc)\tablenotemark{a}} & \colhead{Sp. Type\tablenotemark{a}}& \colhead{Other NUV Obs.}}   
\startdata  
LP 476$-$207 & GJ 3322 & 853 & 12.41 & 6.03 & 6.5 & M4.0Ve & \nodata \\ 
V AD LEO & Gl 388 & 1616 & 10.96 & 3.40 & 4.9 &  M3.0Ve & IUE \\  
LHS 191 & GJ 3289 & 773& 17.16&	2.60&	17.0&	 M6.5V& \nodata \\ 
LHS 3376 & GJ 4053 & 2897 & 14.09 & 2.14 & 7.5 &  M4.5Ve & \nodata \\ 
V 1285 AQL & Gl 735 & 2976 & 10.09 & 2.10 & 10.1 &  M3.0Ve & \nodata\\ 
LP 771$-$95 & GJ 3192 &	538& 11.93& 0.93& 6.4&	 M3.5Ve& \nodata \\ 
LHS 252	& GJ 3512 & 1355& 15.08& 0.31& 9.9& M5.5Ve& \nodata \\ 
V DP DRA & Gl 487 & 2017& 11.28& 0.20& 8.4& M3.0Ve& \nodata \\ 
BD$+$01 2447 & Gl 393 &	1642& 10.32& $-$0.53& 7.3& M2.0V& \nodata \\
HD 42581& Gl 229 & 1021& 9.38& $-$0.51&	5.7& M0.5V& IUE \\ 
HD 199305 & Gl 809 & 3263& 9.20& $-$0.46& 7.4& M0.5V& IUE \\ 
HD 119850 & Gl 526 & 2176& 9.79&$-$0.45& 5.4& M1.5V& \nodata \\ 
HD 165222 & Gl 701 & 2859& 9.85&$-$0.45& 8.1& M1.0V& \nodata \\
BD $+$66 717 & Gl 424 & 1770& 9.56& $-$0.44& 8.9& M0.0V&IUE\\ 
BD $+$11 2576 & Gl 514 & 2123& 9.74& $-$0.43& 7.3& M0.5V& \nodata \\ 
HD 216899 & Gl 880 & 3616& 9.55& $-$0.42& 6.7& M1.5V& \nodata \\
BD $+$70 68b & Gl 48 & 195 & 10.52& $-$0.40& 8.0& M3.0V& \nodata \\ 
HD 36395& Gl 205 & 930&	9.19& $-$0.40& 5.7& M1.5& IUE \\ 
Gl 889.1& \nodata & 3637& 9.76& $-$0.39& 17.0&	 M0.0V& \nodata\\ 
BD $+$18 3421 &Gl 686 & 2792 & 10.14 &$-$0.38 &7.9 & M1.0V& \nodata \\ 
GJ 2066 & \nodata & 1299& 10.38&$-$0.36& 8.8& M2.0V& \nodata \\ 
BD $+$68 946 & Gl 687 & 2797& 10.85& $-$0.34& 4.6& M3.0V& IUE \\ 
BD $+$01 4774 & Gl 908 & 3759& 10.20& $-$0.33& 5.7& M1.0V& IUE \\ 
BD $+$05 1668 & Gl 273& 1168& 11.97& $-$0.30& 3.8& M3.5V& IUE \\ 
BD $+$61 195 & Gl 49 & 199& 9.84& $-$0.28& 8.8& M1.5V& IUE \\
BD $+$44 2051 & Gl 412A & 1723& 10.06& $-$0.26& 5.4& M0.5V& IUE \\ 
V 2306 OPH & Gl 628 & 2599& 12.02& $-$0.23& 4.1& M3.5V& \nodata\\
V IL AQR & Gl 876 & 3604& 11.84& $-$0.20& 4.6& M4.0V& IUE \\
BD $-$40 9712 & Gl 588 & 2446 & 10.45& \nodata& 5.9& M2.5V& \nodata \\
BD $-$44 11909 & Gl 682 & 2779& 12.54& \nodata& 4.8& M3.5V& \nodata \\ 
HD 33793& Gl 191 & 887&	10.56& \nodata& 4.6& M1.0V& IUE \\ 
HD 202560 & Gl 825 & 3326& 8.74&\nodata& 3.9& K7V&	IUE \\ 
HD 191849 & Gl 784 & 3174& 8.99&\nodata& 6.3& K7V&	IUE \\
BD $-$51 5974 & Gl 438 & 1832& 10.36& \nodata& 10.0& M0.0V& \nodata \\ 
\enddata
\tablenotetext{a}{\citet{rhg95}}
\end{deluxetable}

\begin{deluxetable}{lrrrr}
\tablecaption{Measured Fluxes}  
\tablenum{2}  
\tablecolumns{5} 
\tablewidth{0pc}
\tabletypesize{\footnotesize}  
\tablehead{\colhead{Name} & \colhead{F$_{\lambda2385}$} & \colhead{F$_{\lambda2560}$} & \colhead{F$_{\lambda2640}$} & \colhead{F$_{\lambda2800}$} \\ \colhead{} & \multicolumn{4}{c}{$\times$ 10$^{-14}$ ergs s$^{-1}$ cm$^{-2}$}}
\startdata  
GJ 3322&4.57&    2.40&	31.03&     13.77\\
GJ 4053&\nodata&     2.89&0.06&     0.99\\
Gl 735&	18.28&     97.96&18.41&     66.92\\
GJ 3192&3.83&     56.72&0.00&     3.93\\
GJ 3512&0.01&     0.07&	1.35&     \nodata\\
Gl 487&	0.24&     22.20&48.30&     7.26\\
Gl 908&	\nodata&     37.70&82.20&     18.66\\
Gl 229&	228.77&     97.82&540.20&     186.81\\
Gl 809&	135.85&     1391.10&611.09&     108.61\\
Gl 686&	86.84&     \nodata&213.42&     39.94\\
Gl 526&	\nodata&     302.39&202.02&     76.30\\
Gl 701&	252.53&     101.87&145.01&     13.44\\
Gl 424&	\nodata&     \nodata&102.39&     24.87\\
Gl 514&	14.21&     0.004&288.69&     74.54\\
Gl 880&	5.02&     81.50&\nodata	& 86.72\\
Gl 48&	\nodata&     132.56&556.20&     \nodata\\
Gl 205&	181.91&     \nodata&803.12&     222.18\\
GL 889.1&28.52	&     14.53&37.83&     5.48\\
GJ 2066&4.23	&     31.99&59.10&     14.70\\
Gl 687&	\nodata&     184.98&24.50&     \nodata\\
Gl 273&	\nodata&     127.47&50.31&     30.36\\
Gl 49&	33.69&     \nodata&141.92&     32.84\\
Gl 412&	118.46&     113.57&215.92&     27.42 \\
Gl 876&	\nodata&     27.66&92.67&     8.18\\
Gl 588&	115.89&     \nodata&185.24&     \nodata\\
Gl 682&	1.06&    \nodata&55.76&     \nodata\\
Gl 191&	\nodata&     337.53&166.35&     33.22\\
Gl 825&	931.56&     \nodata&2616.64&     201.13\\
Gl 784&	1210.34&     78.76&558.03&     110.58\\
Gl 438&	0.00&     \nodata&59.72	&     21.06\\
Gl 388&	29.05&     175.22&55.45	&    126.99\\
\enddata
\end{deluxetable}

\clearpage


\begin{figure}
\figurenum{1} \epsscale{0.4} \plotone{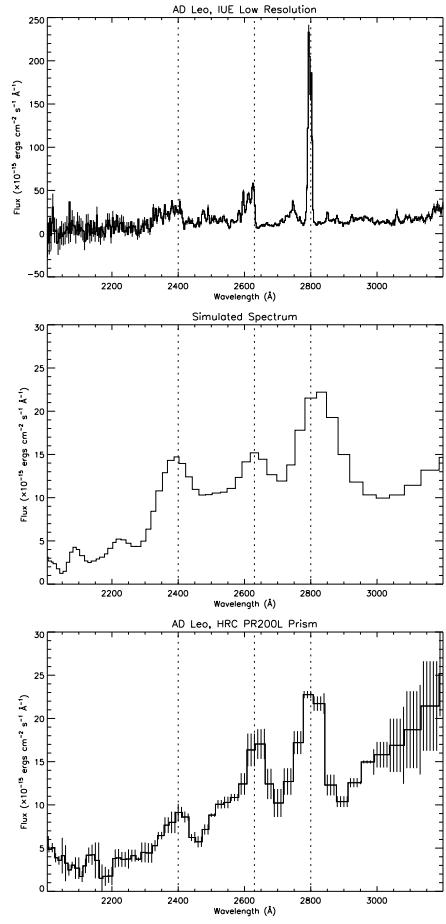}
\caption {AD Leo as observed by IUE (top), prism spectrum simulated from IUE
  using SLIM (middle), and as observed with the HRC PR200L.}
\end{figure}

\clearpage

\begin{figure}
\figurenum{2a} \epsscale{0.8} \plotone{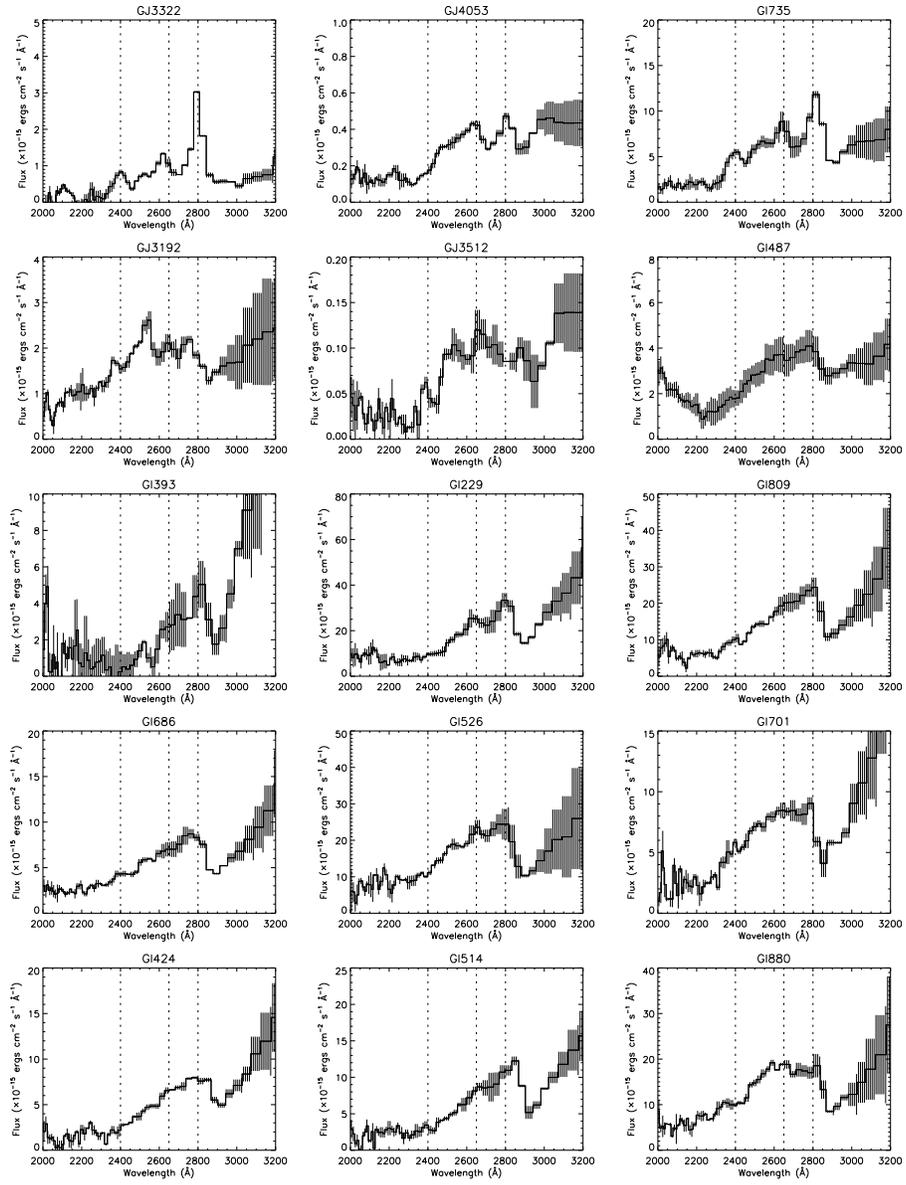}
\caption {ACS PR200L prism spectra.}
\end{figure}
\clearpage
{\plotone{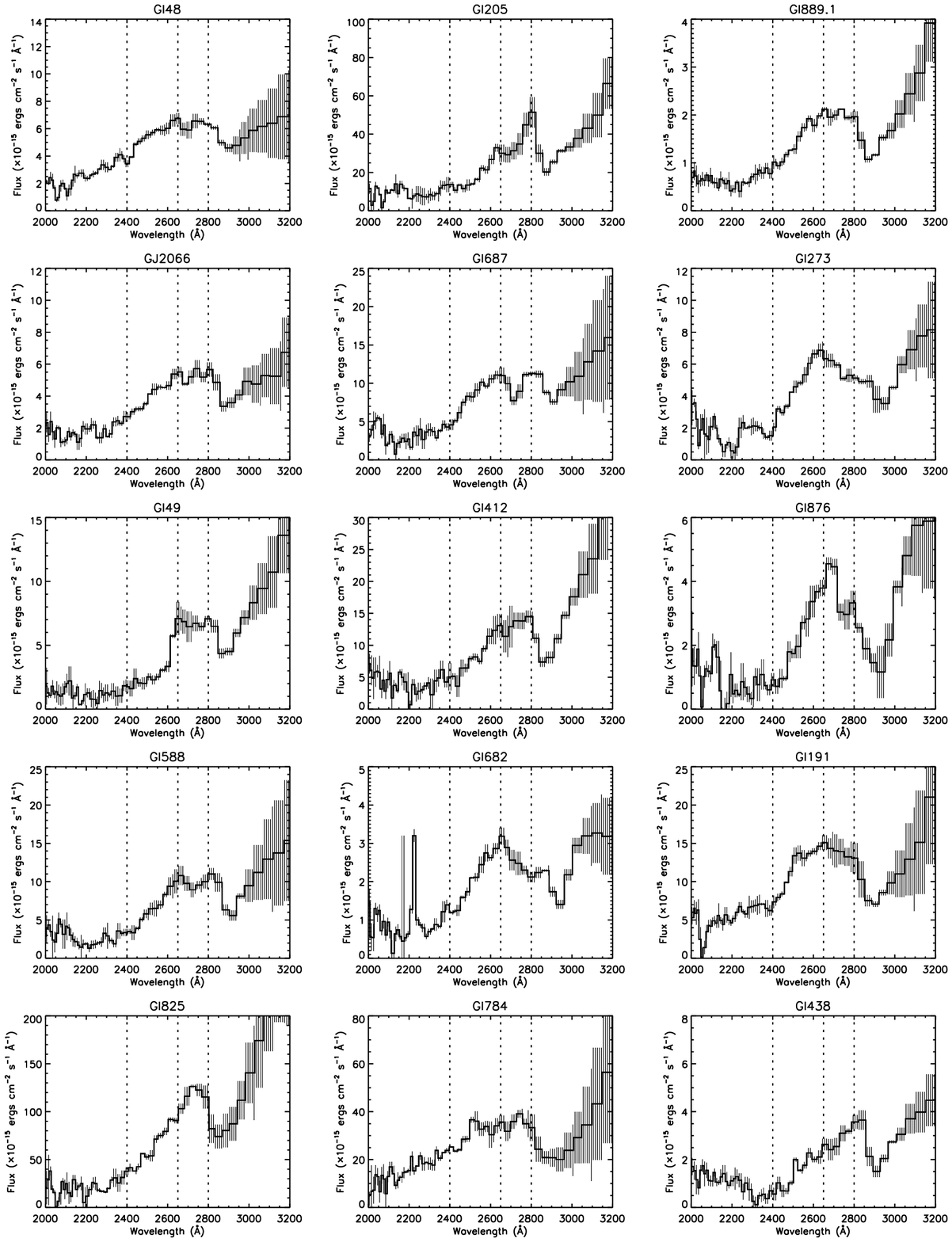}}\\[5mm]
\centerline{Fig. 2b. --- Continued.}

\begin{figure}
\figurenum{3} \epsscale{0.8} \plotone{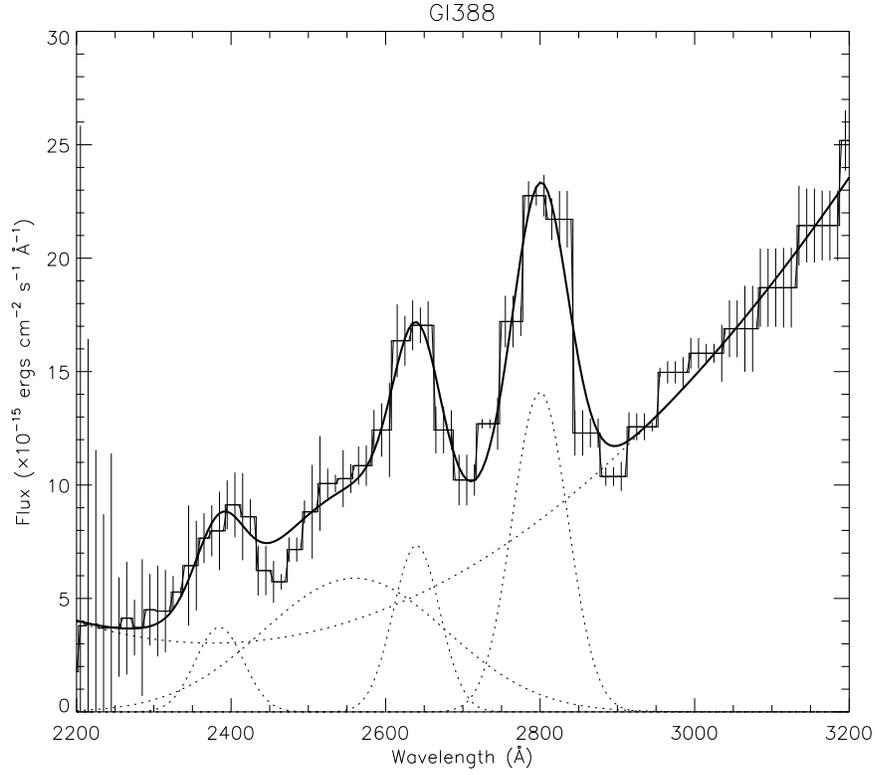}
\caption {Example fit of ACS prism spectra, shown here for AD Leo. The three relatively narrow gaussians represent fits to the clusters of Fe II lines at 2400 and 2600$\mbox{\AA}$ and to the Mg II h and k lines at 2800$\mbox{\AA}$.  The broad gaussian centered near 2500$\mbox{\AA}$ is used
  to fit a collection of unresolved lines between the two main features at 2400
  and 2600$\mbox{\AA}$, together with the slowly varying second-order
  polynomial, which is used to fit the background component of unresolved lines
  and continuum edges throughout the spectrum.}
\end{figure}

\begin{figure}
\figurenum{4} \epsscale{0.8} \plotone{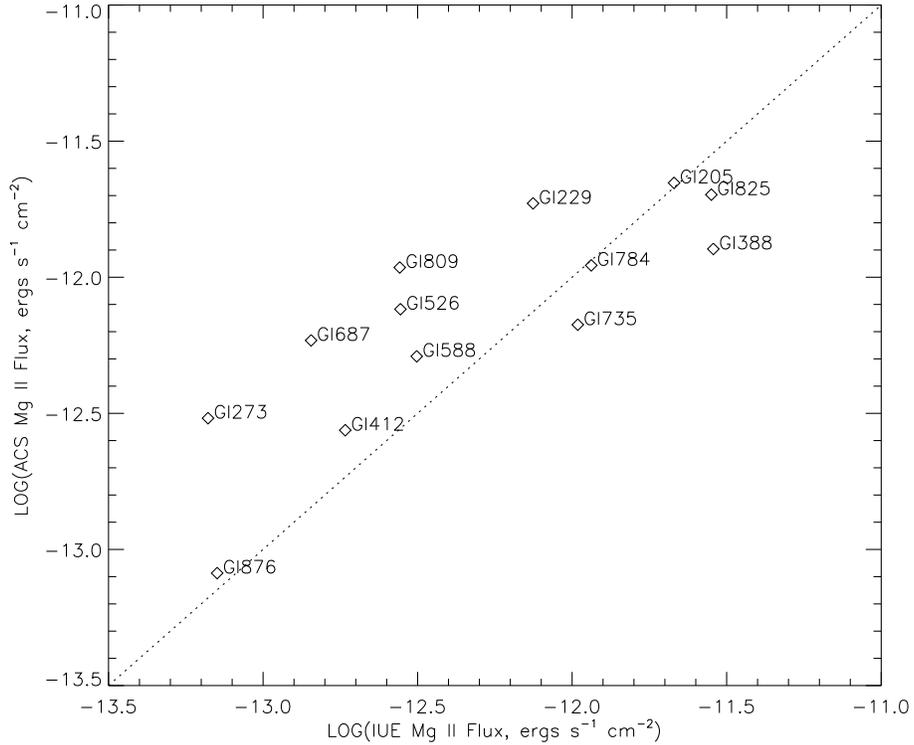}
\caption {Comparison of Mg II surface fluxes measured from ACS and IUE spectra. The dotted line indicates equal fluxes from both instruments. The most active stars (e.g. Gl 388, Gl 735 and Gl 205) tend to live on or below the line, while less active stars show higher ACS fluxes, probably due to contamination from scattered light.}
\end{figure}

\begin{figure}
\figurenum{5} \epsscale{0.8} \plotone{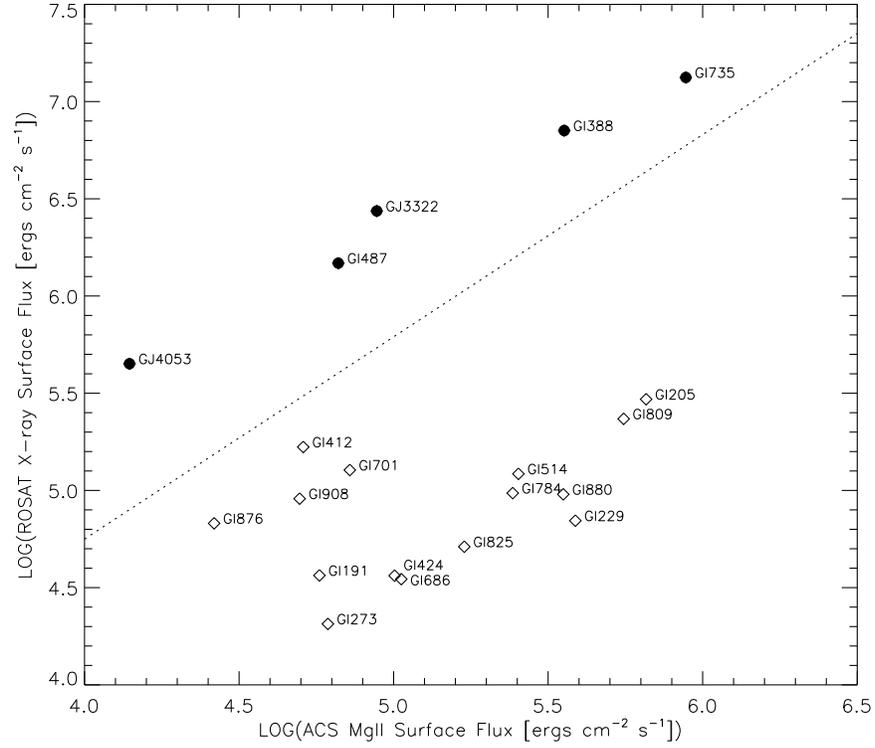}
\caption {Comparison of X ray fluxes from ROSAT with Mg II measured from ACS spectra. The dotted line represents the fit to dMe stars from \citet{md89}. A slight offset exists between their fit and the dMe stars in our sample (plotted as filled circles), as \citet{md89} used X-ray fluxes from Einstein and EXOSAT, while we compare with ROSAT, but otherwise our results are in excellent agreement with the fit.}
\end{figure}

\begin{figure}
\figurenum{6} \epsscale{0.8} \plotone{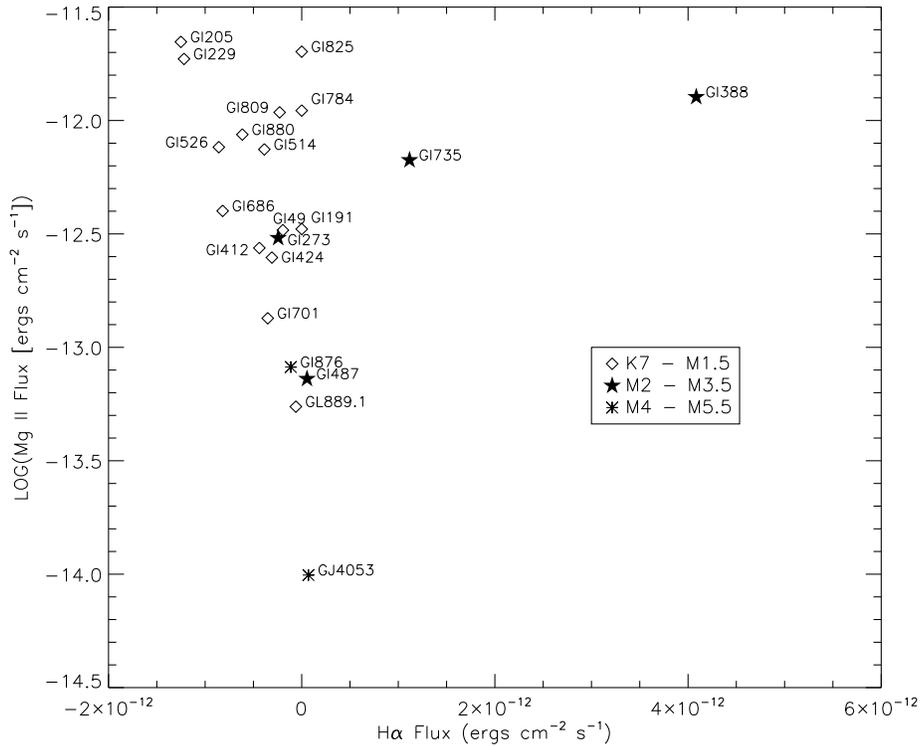}
\caption {Comparison of Mg II measured from ACS spectra with H$\alpha$ fluxes
  calculated using measured H$\alpha$ equivalent widths and the $\chi$ factor
  relation of \citet{wa04}. H$\alpha$ fluxes are shown on a
  linear scale to indicate which are in absorption (negative fluxes) and which
  are in emission (positive flux).}
\end{figure}

\begin{figure}
\figurenum{7} \epsscale{0.8} \plotone{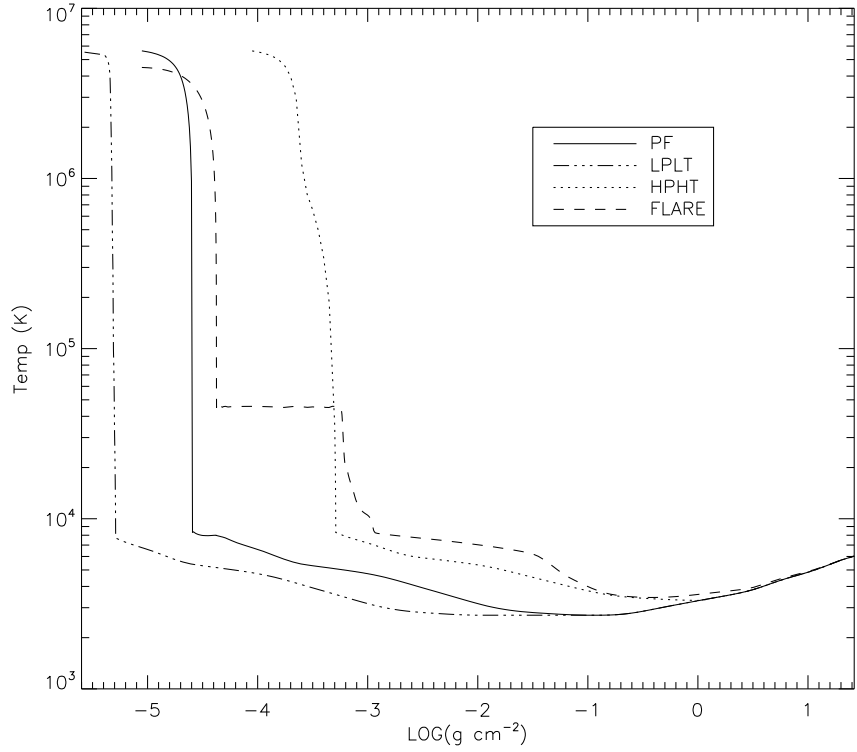}
\caption {Model atmospheres used for Mg II and Fe II calculations. The PF (preflare), LPLT (low temperature low pressure) and HPHT (high pressure high temperature) models all represent quiescent atmospheres at different scalings, while the Flare atmosphere is a snapshot at $\sim$ 85 sec from the hydrodynamic models of \citet{allred05}.}
\end{figure}

\begin{figure}
\figurenum{8} \epsscale{0.8} \plotone{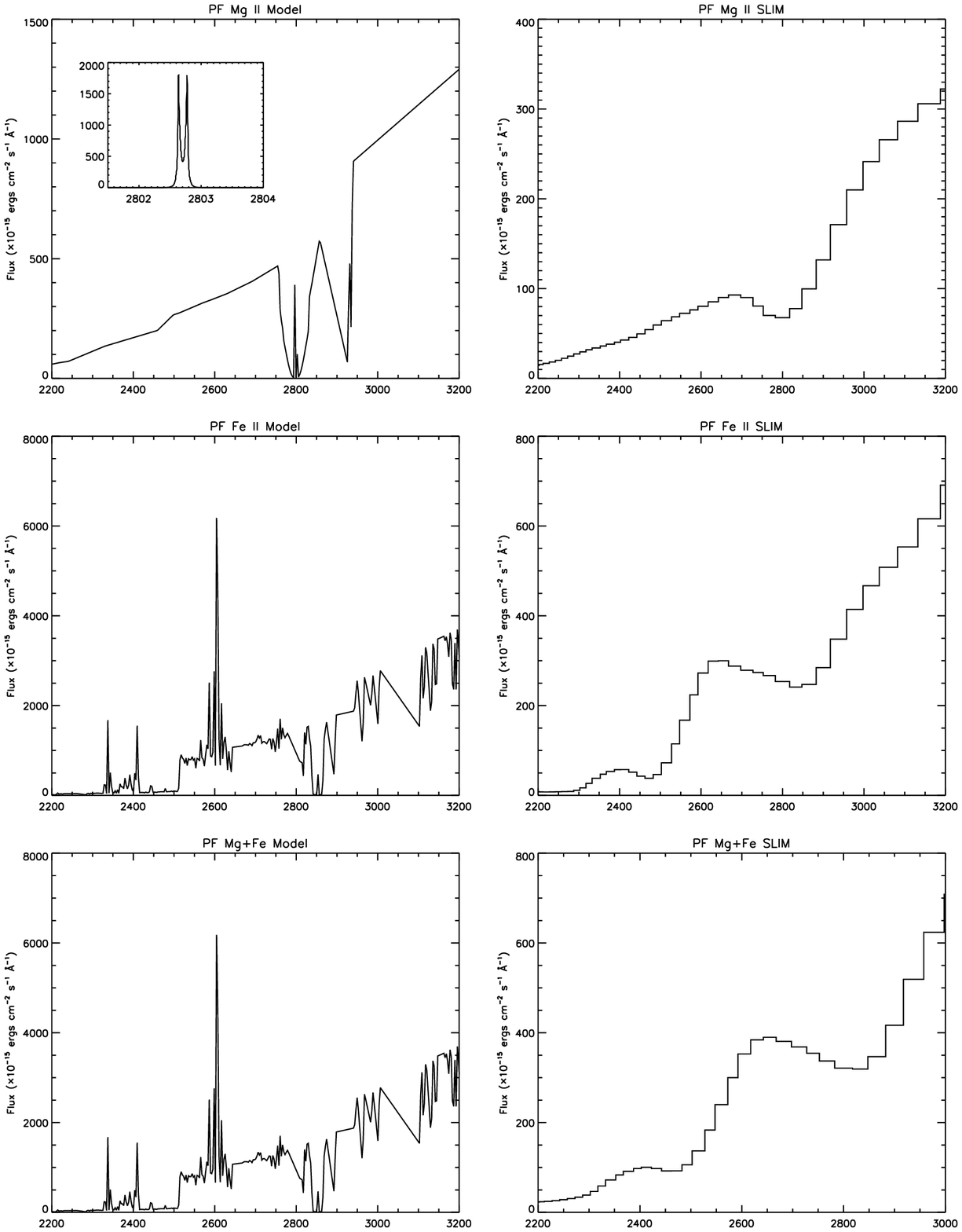}
\caption {Mg II and Fe II Model results for PF atmosphere, degraded to ACS PR200L resolution using SLIM.}
\end{figure}

\begin{figure}
\figurenum{9} \epsscale{0.8} \plotone{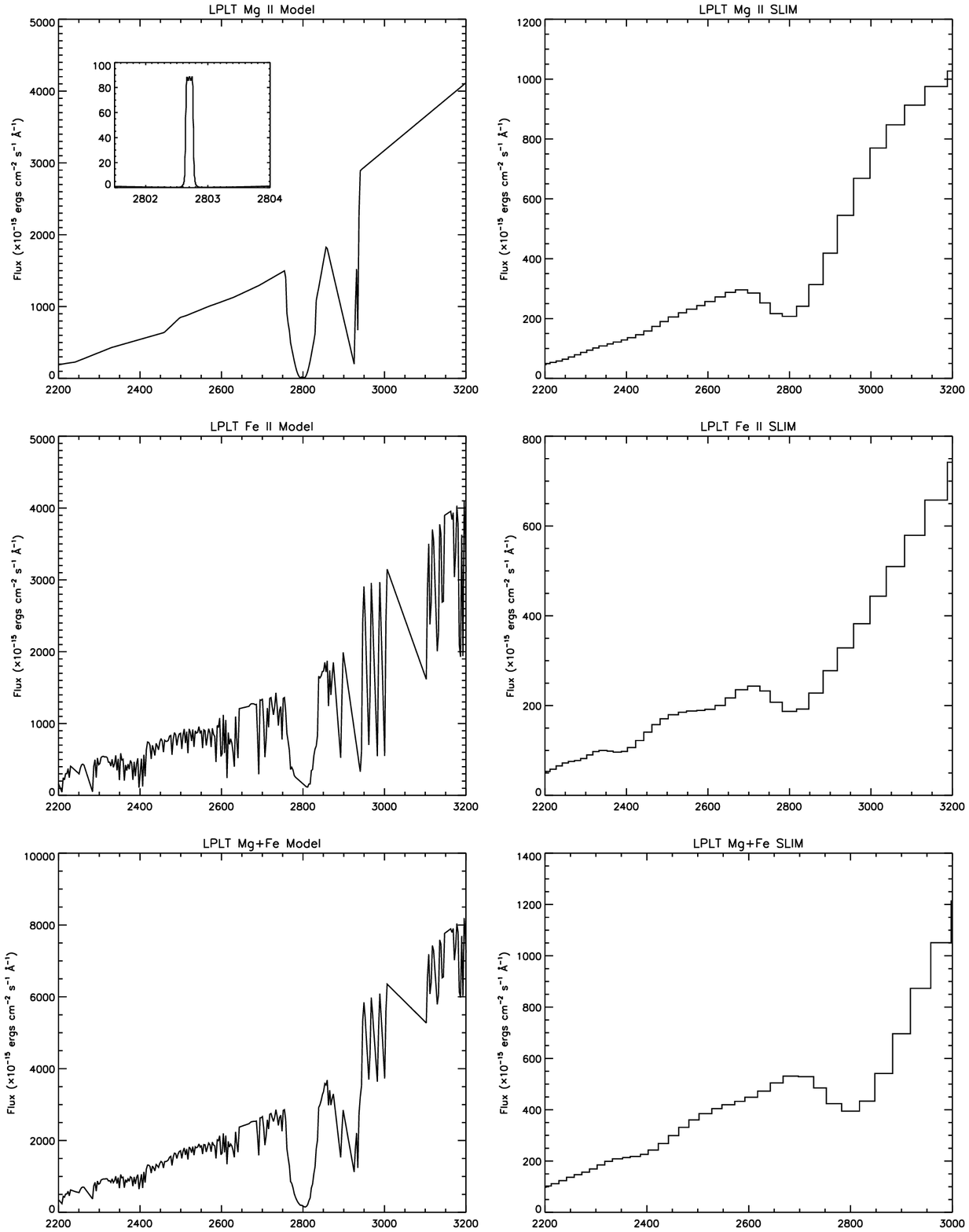}
\caption {Mg II and Fe II Model results for LPLT atmosphere, degraded to ACS PR200L resolution using SLIM.}
\end{figure}

\begin{figure}
\figurenum{10} \epsscale{0.8} \plotone{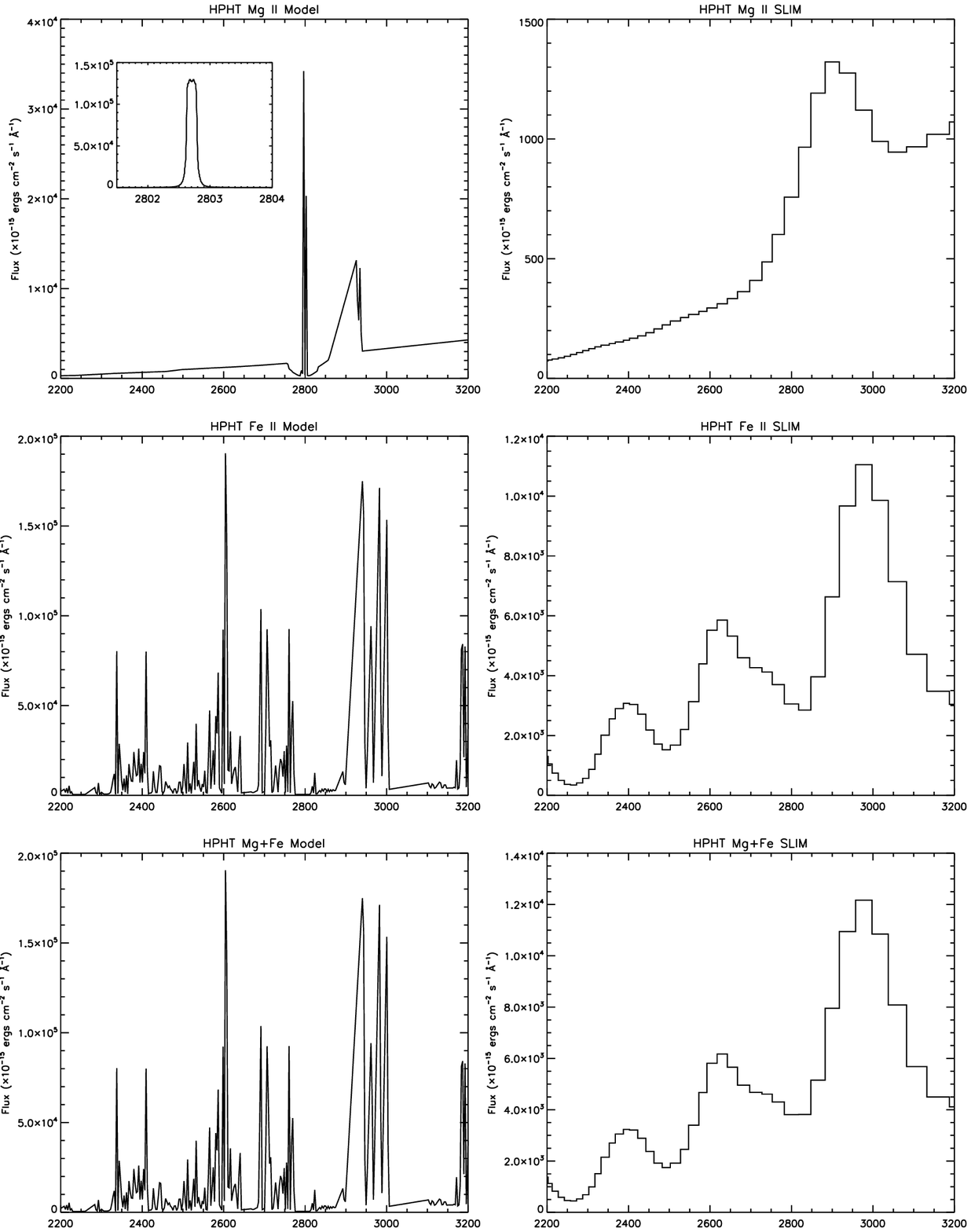}
\caption {Mg II and Fe II Model results for HPHT atmosphere, degraded to ACS PR200L resolution using SLIM.}
\end{figure}

\begin{figure}
\figurenum{11} \epsscale{0.8} \plotone{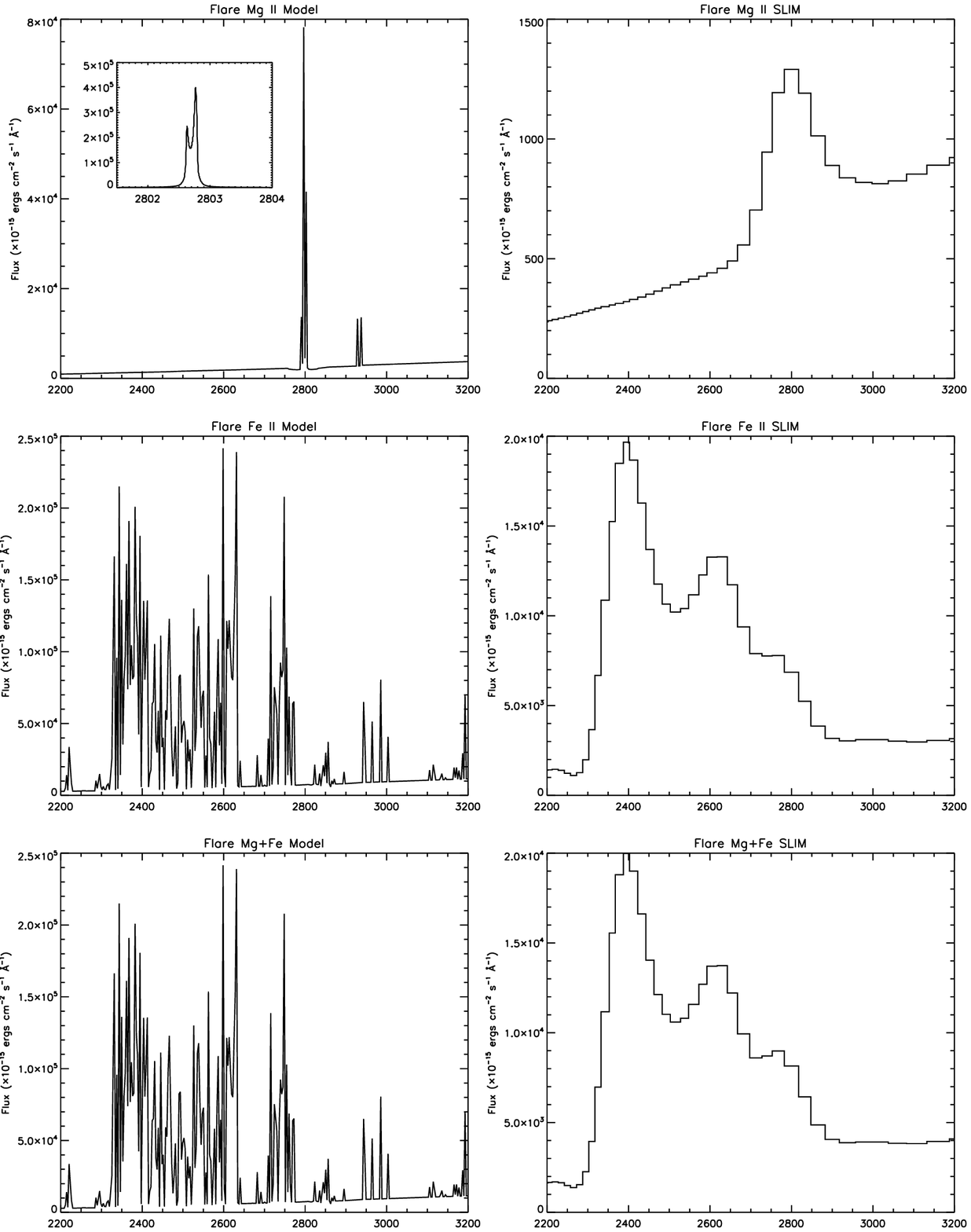}
\caption {Mg II and Fe II Model results for Flare atmosphere, degraded to ACS PR200L resolution using SLIM.}
\end{figure}

\begin{figure}
\figurenum{12} \epsscale{0.8} \plotone{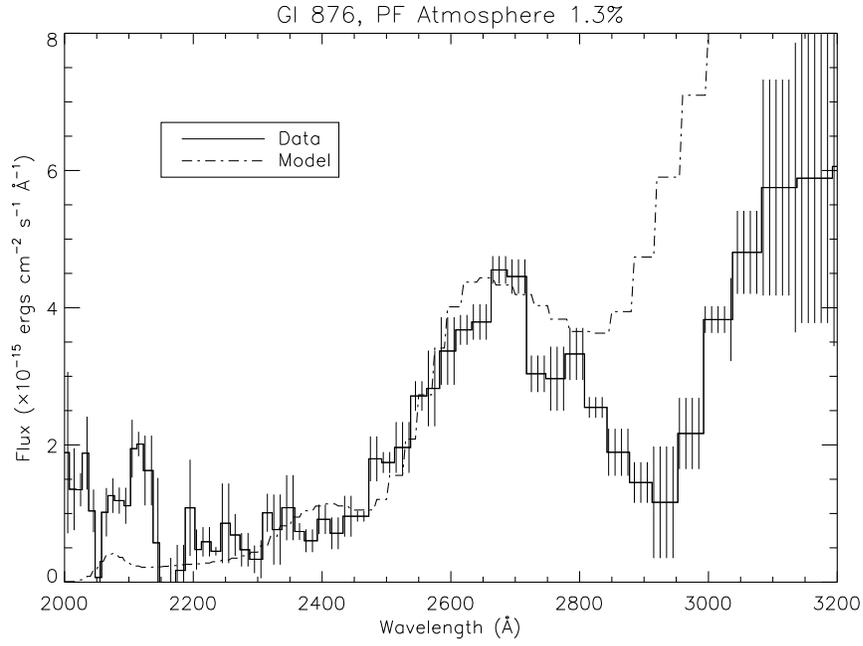}
\caption {Mg II and Fe II Model results for Gl 876. The spectrum is best fit with the ``PF'' atmosphere at a fill factor of 1.3\%. These results show a substantial near-UV flux for even low activity M dwarfs, and are discussed in light of recent habitability models in Section 3.3.1.}
\end{figure}

\begin{figure}
\figurenum{13} \epsscale{0.8} \plottwo{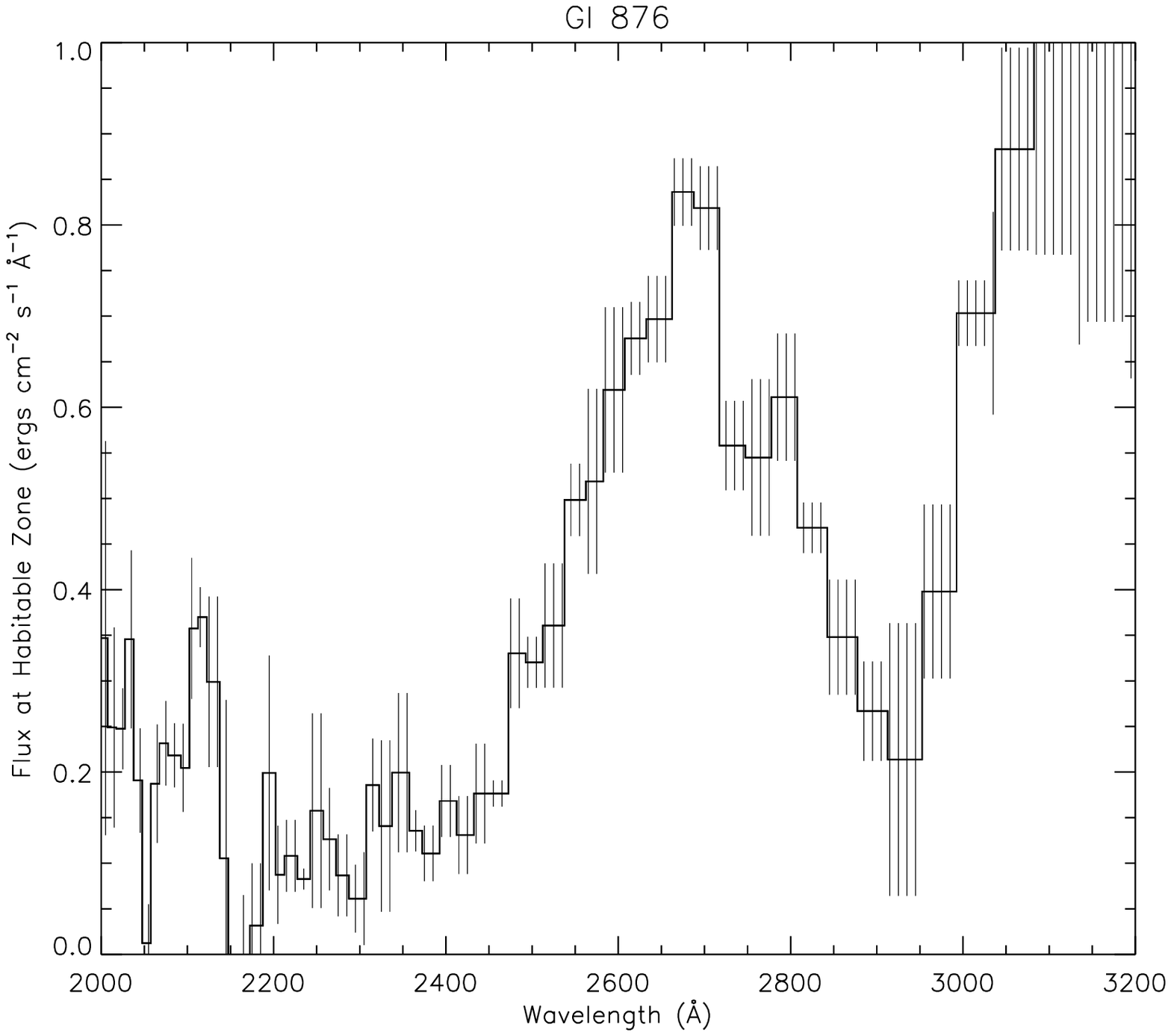}{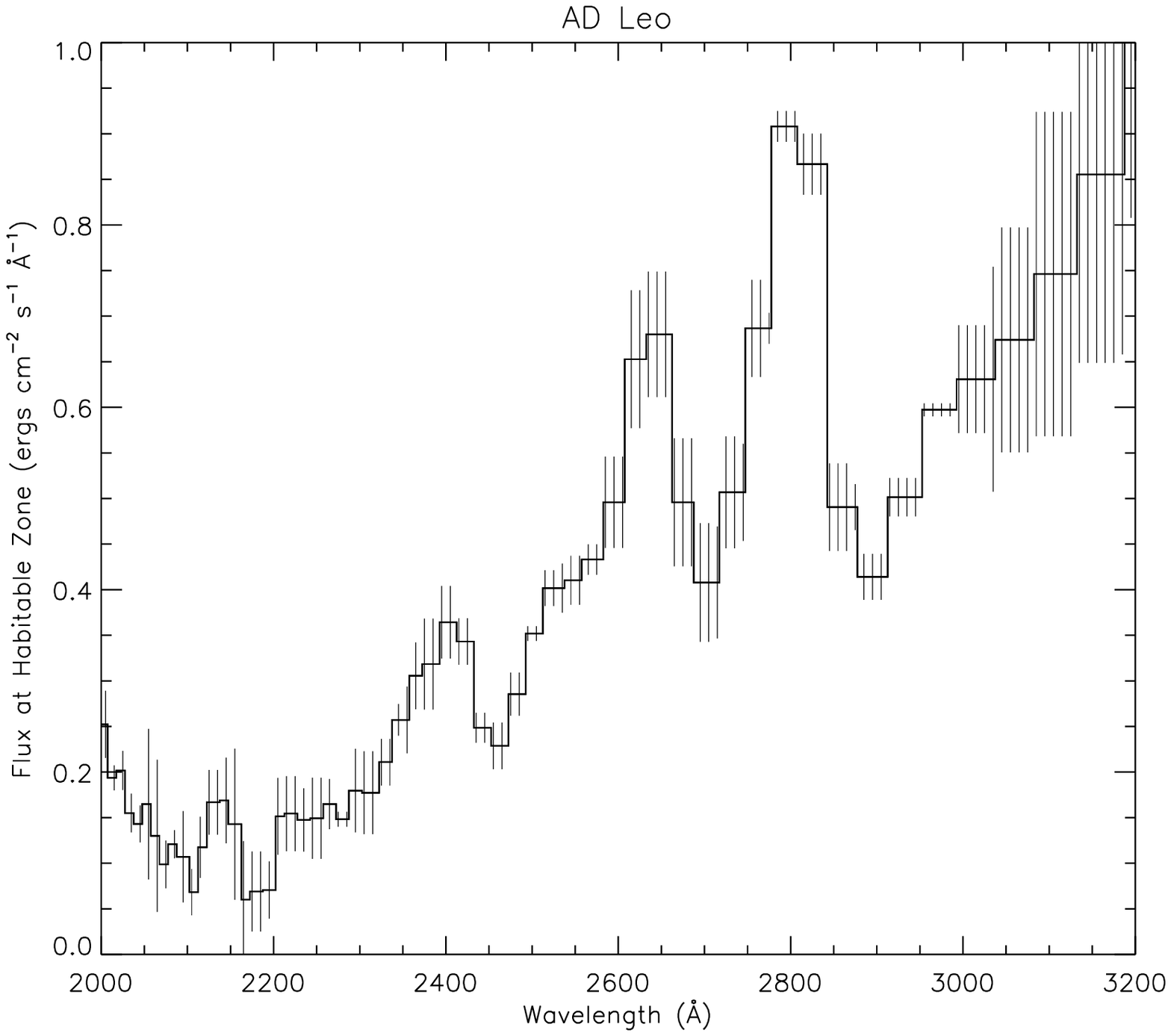}
\caption {Near-UV flux at the habitable zone for Gl 876 and AD Leo. The stellar fluxes are scaled to habitable zone distances given in \citet{seg05}. Although Gl 876 has substantially less activity than AD Leo, its near-UV flux combined with the habitable zone's proximity conspires to make the near-UV top-of-atmosphere flux comparable to AD Leo.}
\end{figure}

\begin{figure}
\figurenum{14} \epsscale{0.8} \plotone{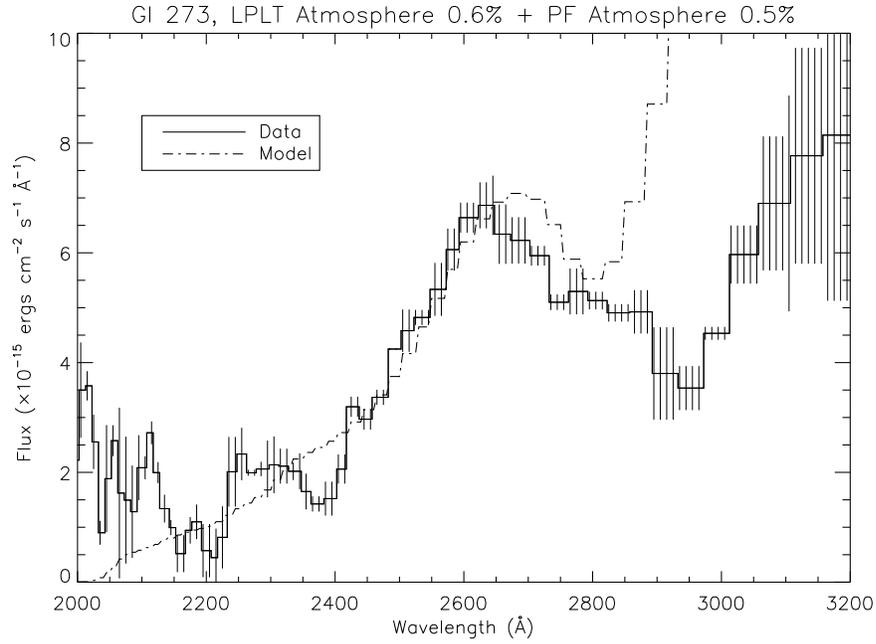}
\caption {Mg II and Fe II Model results for Gl 273. The spectrum is best fit by a combination of LPLT and PF atmospheres in approximately equal portion (0.6\% and 0.5\%).}
\end{figure}

\begin{figure}
\figurenum{15} \epsscale{0.8} \plotone{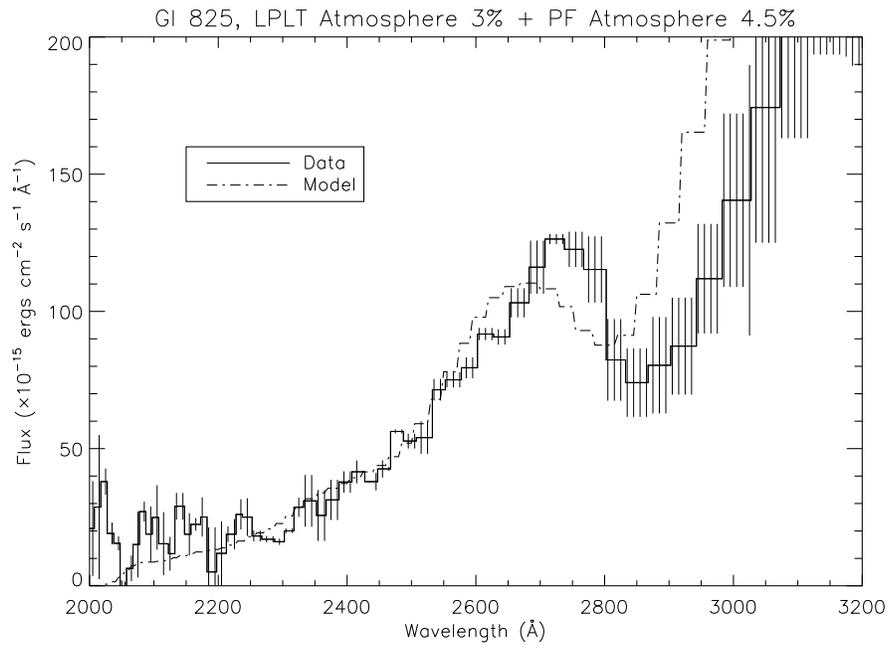}
\caption {Mg II and Fe II Model results for Gl 825. The LPLT and PF atmosphere
  results at fill factors of 3\% and 4.5\%  provide a good fit across the majority of the spectrum, though the peak in flux around 2750$\mbox{\AA}$ is higher than predicted by the models.}
\end{figure}

\begin{figure}
\figurenum{16} \epsscale{0.8} \plotone{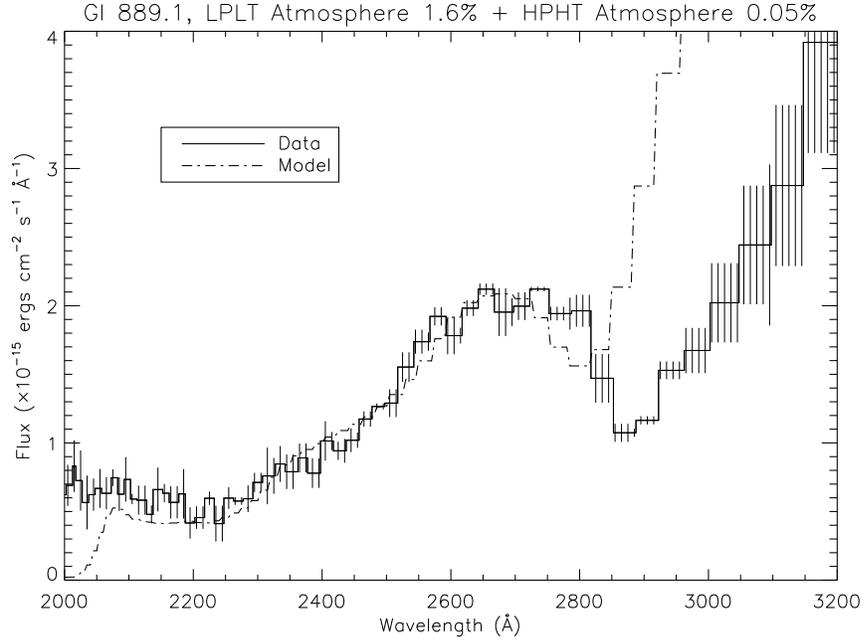}
\caption {Mg II and Fe II Model results for Gl 889.1. The overall shape of the spectrum is reminiscent of Gl 825 and is well fit by the LPLT atmosphere (1.6\%), but the hotter HPHT component must also be included (0.05\% to match the short wavelength flux from 2000 to 2200$\mbox{\AA}$.}
\end{figure}

\begin{figure}
\figurenum{17} \epsscale{0.8} \plottwo{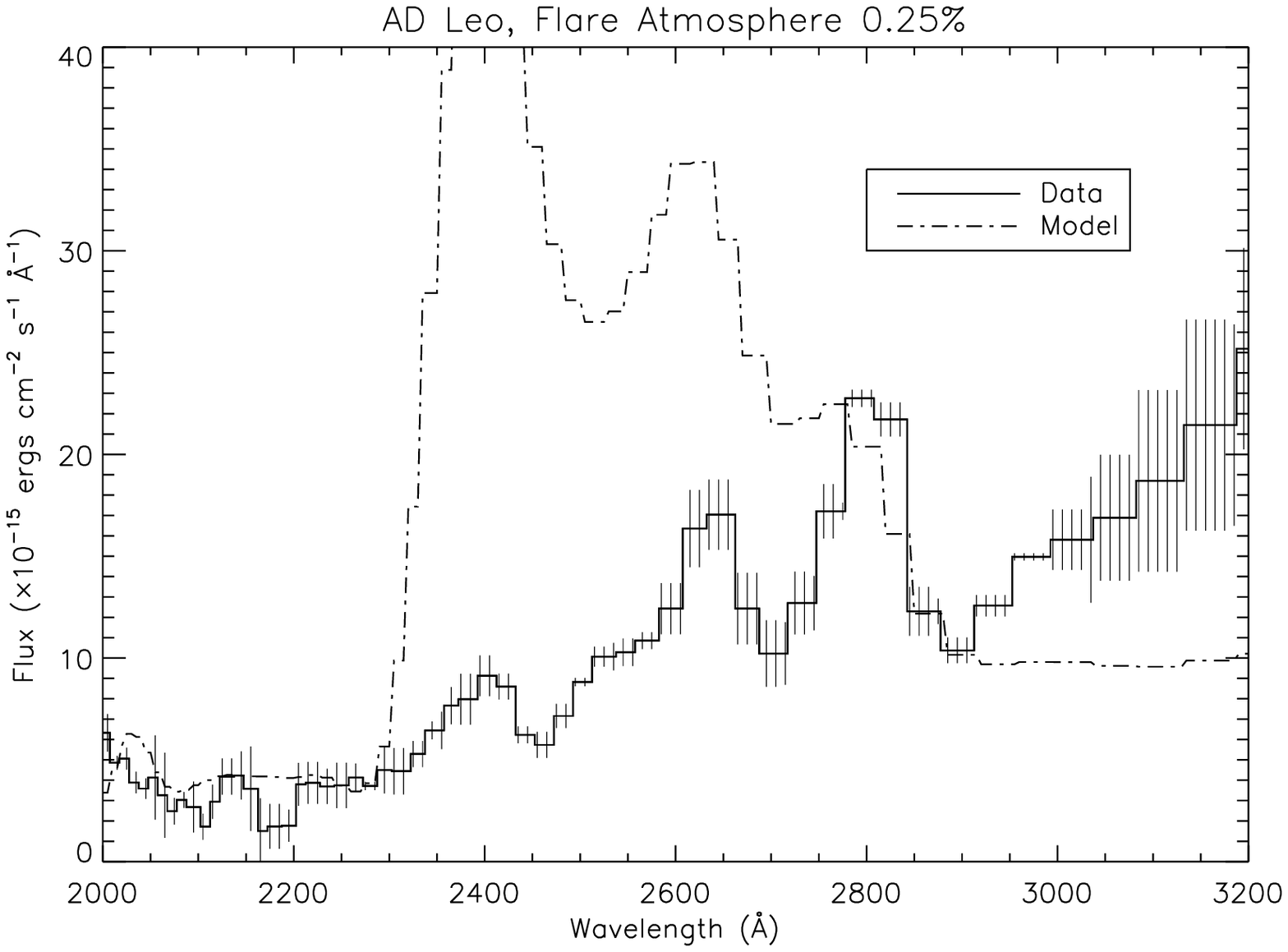}{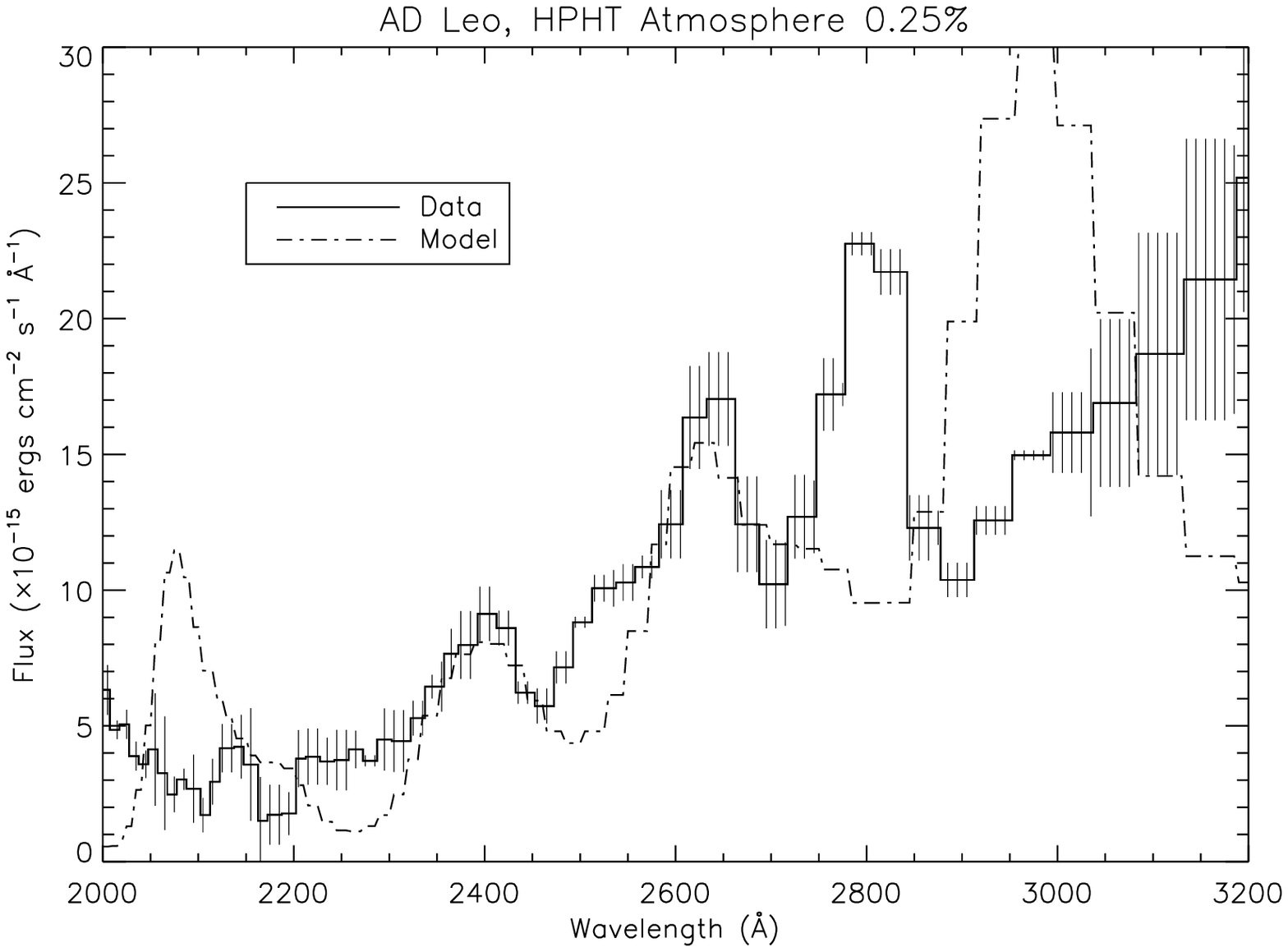}
\caption {Mg II and Fe II Model results overplotted with AD Leo data-- left: Flare atmosphere, right: HPHT atmosphere. Our hottest ``Flare'' model, with a fill factor of 0.25\%, roughly matches the flux around 2800$\mbox{\AA}$ and at wavelengths shorter than 2200$\mbox{\AA}$, but the Fe II emission at $\sim$ 2400$\mbox{\AA}$ and $\sim$2600$\mbox{\AA}$ is vastly overestimated. The HPHT atmosphere (also at a fill factor of 0.25\%) produces a much better match to these two Fe II features, but unfortunately does not account for any of the observed Mg II flux.}
\end{figure}

\end{document}